\documentclass[preprintnumbers, prl, twocolumn, showpacs, floatfix, preprintnumbers, letterpaper, nofootinbib, amsmath, amssymb, superscriptaddress]{revtex4}
\usepackage{amssymb,amsmath,bm,dcolumn,epsf,graphicx,latexsym,slashed,simplewick}
\usepackage{color}
\usepackage[colorlinks,linkcolor=magenta,anchorcolor=blue,citecolor=blue]{hyperref}

\def\be{\begin{equation}}
\def\ee{\end{equation}}
\def\bea{\begin{eqnarray}}
\def\eea{\end{eqnarray}}

\begin{document}

\title{Perturbations of bounce inflation scenario from $f(T)$ modified gravity revisited}

\author{Taotao Qiu}
\email{qiutt@mail.ccnu.edu.cn}
\affiliation{Key Laboratory of Quark and Lepton Physics (MOE) and College of Physical Science $\&$ Technology, Central China Normal University, Wuhan 430079, P.R.China}

\author{Kun Tian}
\email{tiankun@mails.ccnu.edu.cn}
\affiliation{Key Laboratory of Quark and Lepton Physics (MOE) and College of Physical Science $\&$ Technology, Central China Normal University, Wuhan 430079, P.R.China}

\author{Shaojun Bu}
\email{bushaojun@mails.ccnu.edu.cn}
\affiliation{Key Laboratory of Quark and Lepton Physics (MOE) and College of Physical Science $\&$ Technology, Central China Normal University, Wuhan 430079, P.R.China}


\begin{abstract}
In this work, we revisit the perturbations that are generated in the bounce inflation scenario constructed within the framework of $f(T)$ theory. It has been well known that pure $f(T)$ theory cannot give rise to bounce inflation behavior, so aside from the gravity part, we also employ a canonical scalar field for minimal extension. We calculate the perturbations in $f(T)$ theory using the well-established ADM formalism, and find various conditions to avoid their pathologies. We find that it is indeed very difficult to obtain a healthy model without those pathologies, however, one may find a way out if a potential requirement, say, to keep every function continuous, is abandoned.
\end{abstract}
\maketitle

{\it Introduction.}
Inflation \cite{Guth:1980zm, Linde:1981mu, Starobinsky:1980te} has been viewed as one of the most successful theory in modern cosmology. Not only is it simple and elegant, it can also simultaneously solve several Big-Bang puzzles, as well as predicts nearly scale-invariant power spectrum, which is verified by the observational data \cite{Akrami:2018odb}. Nonetheless, inflation cannot avoid the notorious Big-Bang Singularity, whose existence has been proved by S. Hawking and R. Penrose \cite{Hawking:1969sw, Borde:1993xh}. one of the easiest ways of avoiding the singularity point might be to assume that the universe starts from a contracting phase, and bounce into the expanding one as we observed \cite{Novello:2008ra}. Together with the inflation period that follows up, this can be called as ``bounce inflation" scenario of the early universe \cite{Piao:2003zm}.  

For the universe to bounce, some conditions must be satisfied, such as the Null Energy Condition violating \cite{Cai:2007qw}. In order to do so, one may either introduce exotic matter which can violate the NEC, or modify the classical General Relativity. Recently the studies of bounce cosmology encountered a boost in the literatures and fruitful bounce models are built in both two ways. The first way includes double-scalar-field bounce \cite{Cai:2007zv} and higher-order single-scalar-field bounce \cite{Qiu:2011cy}, while the second way includes nonminimal coupling bounce \cite{Qiu:2010ch}, $f(R)$ bounce \cite{Carloni:2005ii}, $f(T)$ bounce \cite{Cai:2011tc}, Loop-quantum bounce \cite{Date:2004fj} and so on. 

In Ref. \cite{Libanov:2016kfc} (see also \cite{Kobayashi:2016xpl, Ijjas:2016tpn}), it is proved that it is indeed very difficult for a single scalar to make a alhealthy bounce (inflation) scenarios, which needs to go even beyond Horndeski theory \cite{Cai:2016thi}. However, the conclusion only applies to single scalar models, and for modified gravity driven bounces whether there is such ``no-go" theorem is unknown. In this letter, we will focus on an interesting type of bounce inflation scenario, driven by the $f(T)$ modified gravity theory. The $f(T)$ theory is an extension of the so-called ``Teleparallel Equivalent General Relativity (TEGR)". Although TEGR, as a torsion theory, is equivalent to General Relativity,  $f(T)$ is no longer equivalent to the extension to GR, namely $f(R)$ theory, but act as a totally new theory, with many interesting properties not shared by GR or $f(R)$ theories. For more information on $f(T)$ theory, see reviews \cite{Hehl:1976kj}. In the following, we will perform a detailed investigation of perturbations generated by $f(T)$ modified gravity theory, and apply it into the bounce inflation scenario. Moreover, we study on what conditions could the perturbations remain healthy passing through the bounce. 

{\it $f(T)$ modified gravity and the bounce inflation ansatz.}
We start with the general action of $f(T)$ modified gravity \cite{Hehl:1976kj}:
\be
\label{action}
S=\int d^4x\left[ef(T)+\sqrt{-g}{\cal L}_m\right]~,
\ee
where $ef(T)$ is the gravity part, while ${\cal L}_m$ is the matter part which could be added to the gravity, and   $T$ is the torsion scalar constructed from the torsion tensor:
\be
\label{torsionscalar}
T\equiv\frac{1}{2}T^\rho_{~\mu\nu}(\delta^\mu_\rho T^{\alpha\nu}_{~~\alpha}-\delta^\nu_\rho T^{\alpha\mu}_{~~\alpha})-\frac{1}{4}T^\rho_{~\mu\nu}(T^{\mu\nu}_{~~\rho}-T^{\nu\mu}_{~~\rho}-T^{~\mu\nu}_{\rho})~,
\ee
while torsion tensor is defined as the antisymmetric part of the affine connection $\Gamma^{\rho}_{~\mu\nu}$:
\be
\label{torsiontensor}
T^{\rho}_{~\mu\nu}\equiv\Gamma^{\rho}_{~\nu\mu}-\Gamma^{\rho}_{~\mu\nu}~.
\ee
Note that in non-Riemannian geometry, the affine connection is not necessarily symmetric. Actually, with the existence of torsion tensor, the affine connection $\Gamma^{\rho}_{\mu\nu}$ can no longer be expressed in terms of metric, but act as an independent variable \cite{Hehl:1976kj}. 

%\be
%\Gamma^{\rho}_{~\mu\nu}=\tilde{\Gamma}^{\rho}_{~\mu\nu}-K^{\rho}_{~\mu\nu}~,
%\ee
%where $\tilde{\Gamma}^{\rho}_{~\mu\nu}$ is the Christoffel symbol which can be expressed in terms of metric $g_{\mu\nu}$, namely 
%and
%\be
%\label{contorsion}
%K^{\mu\nu}_{~~\rho}\equiv-\frac{1}{2}(T^{\mu\nu}_{~~\rho}-T^{\nu\mu}_{~~\rho}-T^{~\mu\nu}_{\rho})~,
%\ee
%is the so-called contorsion tensor. Furthermore, the Ricci scalar $R$ can be calculated as:
%\bea
%R&\equiv&g^{\mu\nu}(\partial_\alpha\Gamma^\alpha_{~\mu\nu}-\partial_\nu\Gamma^\alpha_{~\mu\alpha}+\Gamma^\alpha_{~\mu\nu}\Gamma^\beta_{~\alpha\beta}-\Gamma^\alpha_{~\mu\beta}\Gamma^\beta_{~\mu\alpha})~\nonumber\\
%&=&\tilde{R}+\Delta R~,
%\eea
%where $\tilde{R}$ is the Ricci scalar made of Christoffel symbols, and 
%\be
%\Delta R=K_{\rho\lambda\nu}K^{\nu\rho\lambda}-4\nabla_\mu T^\mu-4T_\mu T^\mu~
%\ee
%is the difference between $R$ and $\tilde{R}$. In the above formula, $T_\mu\equiv T^{~~\nu}_{\mu\nu}$ and $\nabla_\mu T^\mu\equiv\partial_\mu T^\mu-\Gamma^\mu_{~\mu\nu}T^\nu$ is the covariant derivative of $T^\mu$. 
%
%If there is no torsion, say, $\Gamma^{\rho}_{~\mu\nu}\rightarrow\tilde{\Gamma}^{\rho}_{~\mu\nu}$ which is symmetric, then $\Delta R=0$ and $R$ reduces to $\tilde R$ as in GR. 
In $f(T)$ theory there will be no curvature. In order to do so, the connection is chosen as $\Gamma^{\rho}_{~\mu\nu}=e^\rho_A\partial_\nu e^A_\mu$ (Weitzenbock connection \cite{Aldrovandi:1996ke}) where $e^A_\mu$ is the tetrad with an internal index $A=0,1,2,3$. A relation between tetrad and normal metric $g_{\mu\nu}$ can be given as $e^A_\mu e^B_\nu \eta_{AB}=g_{\mu\nu}$, $e\equiv|e^A_\mu|=\sqrt{-g}$ via a flat metric $\eta_{AB}=diag(-1,1,1,1)$. With the relation, the Weitzenbock connection can be related to the Christoffel symbol in GR, $\tilde{\Gamma}^{\rho}_{~\mu\nu}=g^{\rho\alpha}(\partial_\nu g_{\mu\alpha}+\partial_\mu g_{\nu\alpha}-\partial_\alpha g_{\mu\nu})/2$, as $\Gamma^{\rho}_{~\mu\nu}=\tilde{\Gamma}^{\rho}_{~\mu\nu}-K^{\rho}_{~\mu\nu}$ where the contorsion tensor $K^{\mu\nu}_{~~\rho}\equiv(T^{\nu\mu}_{~~\rho}+T^{~\mu\nu}_{\rho}-T^{\mu\nu}_{~~\rho})/2$. Moreover, the torsion scalar $T$ related to the Ricci scalar $\tilde R$ in GR as:
\be
\label{TandR}
T=\tilde{R}+2\tilde{\nabla}^\nu T^\mu_{~\mu\nu}~,
\ee
which shows that $T$ and $\tilde R$ only differs by a total derivative, therefore an action containing only $T$ (TEGR) as its Lagrangian is actually nothing but GR \cite{Aldrovandi:2013wha}.

According to the action (\ref{action}), the Friedmann equations in flat FRW spacetime ($e^A_\mu=diag(1,a(t),a(t),a(t))$) turn out to be:
\be
\label{Friedmann}
H^2=\frac{1}{2f_T}\left(\frac{8\pi G}{3}\rho_m-\frac{f(T)}{6}\right)~,~\dot{H}=-\frac{4\pi G(\rho_m+p_m)}{f_T-12H^2f_{TT}}~,
\ee
where $\rho_m$ and $p_m$ are energy density and pressure coming from ${\cal L}_m$, and $_{,T}$ denotes derivative with respect to $T$. Also note that $T=-6H^2$ in flat FRW spacetime. Eq. (\ref{Friedmann}) can be deformed into:  
\be
\dot f(t)=f_{,T}\dot T(t)=-6\frac{\dot H}{H}\left[\frac{8\pi G}{3}\rho_m(t)-\frac{f[T(t)]}{6}\right]~,
\ee
which has a general analytical solution:
\be
\label{ft}
f(t)=e^{-\int{P(t)}dt}\left[\lambda+\int{Q(t)e^{\int{P(t)}dt}}\right]~,
\ee
with $P(t)=-\dot H/H$, $Q(t)=-16\pi G\rho_m(t)\dot H/H$, and $\lambda$ is the integration constant. 

We are focusing on the bounce inflation solution given by $f(T)$ modified gravity theory. The bounce, by definition, is the scenario where the universe goes from contracting phase ($H<0$) to expanding phase ($H>0$), therefore there must be a pivot point where $H=0$, $\dot H>0$ is satisfied, which we call the bounce point. However, in absence of the matter part, namely $\rho_m=p_m=0$, from Eqs. (\ref{Friedmann}) one can only get a trivial solution of $f(T)=-T/3+\lambda/\sqrt{-T}$ with an integral constant $\lambda$, and $\dot H=0$ forever, so no bounce will happen. This is a well-known result \cite{Cai:2011tc} and that's why a matter part will be needed. Moreover, in order to solve the inconsistency problem in usual bounce model with single scalar degree of freedom (namely one cannot both solve the anisotropy problem and get the scale-invariant power spectrum) (last two references in \cite{Qiu:2011cy}), we explore the bounce inflation model where the contracting phase has a large equation of state, $w\geq 1$, or, in terms of the slow-varying parameter $\epsilon\equiv 3(1+w)/2$, $\epsilon\geq 3$, while in expanding phase usual slow-roll conditions for inflation, $w\simeq -1$, $\epsilon\simeq 0$, is imposed. Note that other bounce inflation solutions in $f(T)$ theory has been discussed in Ref. \cite{Bamba:2016gbu}.

In principle, one can employ the reconstruction method to obtain the functional form of $f(T)$, which gives the bounce inflation solution, as has been done in \cite{Cai:2011tc, Nojiri:2017ncd}. However, there will be several conditions coming from perturbations, namely ghost-free and gradient stable conditions for both scalar and tensor perturbations, violating any of which will make the model pathologic. So before heading to specific models, we will first analyze the perturbation theory of $f(T)$ in a very general form, to find whether these conditions will impose rigid constraints on $f(T)$ models.  

{\it Perturbations generated from $f(T)$ modified gravity.}
\label{perturbation}
In order to calculate the perturbations in $f(T)$ bounce inflation scenario, first of all we write down the tetrads containing perturbation as:
\bea
&&e^0_{~\mu}=(N,\tilde{N}_i)~,~e^a_{~\mu}=(N^a,h^a_{~i})~,\nonumber\\
&&e^{~\mu}_0=\left(\frac{1}{N},-\frac{N^i}{N}\right)~,~e^{~\mu}_a=(0,h^{~i}_a)~,
\eea
where $a=1,2,3$ is the spatial part of internal indices, $N$ is the lapse function, $N^a$ is the shift vector, and $h^a_i$ is the induced 3-vierbein. Note that although the metric is symmetric, the tetrad used to construct it does not need to be symmetric, therefore $\tilde{N}_i$ and $N^a$ has independent components. However, both $\tilde{N}_i$ and $N^a$ can be decomposed into a pure vector and gradient of a scalar, say, $\tilde{N}_i=\partial_i\beta+u_i$, $N^a=\delta^a_{~i}(\partial_i B+v_i)$ \cite{Izumi:2012qj}. In this paper we don't consider vector perturbations, and to make the calculations simpler, hereafter we set $\beta=0$ as a gauge fixing. By making use of the relation $e^A_\mu e^B_\nu \eta_{AB}=g_{\mu\nu}$, we can get the line element as: $ds^2=N^2dt^2-h_{ij}(dx^i+N^idt)(dx^j+N^jdt)$, and $e=\sqrt{-g}=\sqrt{h}N$, which is consistent in the result of ADM metric decomposition usually used in Riemannian gravity theories. 

Besides the gravity part, in principle the matter part can also have perturbations. However, since we mainly focus on the perturbations generated in $f(T)$ gravity, we for simplicity turn off the perturbations for matter part, as is valid if the (isocurvature) perturbations generated by matter is quite small. As an explicit example, we set it to be a canonical scalar field:
\be
{\cal L}_m=\frac{1}{2}(\nabla\phi)^2-V(\phi)~.
\ee

Perturbing the tetrad as: $h^{a}_{~i}=a(\delta^{a}_{~i}+\frac{1}{2}\gamma^{a} _{~i})$, and after tedious calculations, we obtain the second order tensor perturbation action as:
\be
\delta^T_2S\subset\frac{1}{8}\int dtd^{3}xa^{3}f_{T}\big|_0(\dot{\gamma_{ij}}\dot{\gamma^{ij}}-a^{-2}\gamma^{ij}_{,k}\gamma_{ij}^{,k})~.
\ee
From this action one can see that it is very much alike that of GR, except for the coefficients in front of both kinetic term and spatial derivative terms are $f_T\big|_0$, with the sound speed squared being unity. If $f_{T}=1$, we can get back to GR as it must be. For $f_{T}\neq 1$ case, in order for the tensor perturbation to be free of both ghost and gradient instabilities, one should require $f_{T}$ be positive definite. Therefore the first condition to have healthy perturbation is:\\
\textbf{1) from stability of tensor perturbation:}\\
\be
\label{con1}
f_T>0~.
\ee

Similarly, perturbing the tetrad as $N=1+\alpha$, $N_i=\partial_i{\psi}$, and $h_{ij}=a^2e^{2\zeta}\delta_{ij}$ ($\alpha$ and $\psi$ are non-dynamical variables), one gets the second order scalar perturbation action as:
%\eeWith Eq.s (\ref{perturb}), (\ref{constraint1}) and (\ref{constraint2}), one can perform straightforward calculation to obtain:
\be
\label{pertaction}
\delta^{(2)}S=\int d^4x[\alpha_1\zeta^{\prime 2}-\alpha_2(\partial\zeta)^2-\alpha_3(\partial^2\zeta)^2]~,
\ee
where ``$^\prime$'' denotes derivative with respect to conformal time, $\eta\equiv\int a^{-1}(t)dt$, and 
\bea
\label{alpha1}
\alpha_1&=&-\frac{a\dot{\phi}^2f_T}{H(\dot{f_T}-3Hf_T)}~,\\
\label{alpha2}
\alpha_2&=&-2af_T-\frac{d}{dt}\left(\frac{6af_T^2}{\dot{f_T}-3Hf_T}\right)~,\\
\label{alpha3}
\alpha_3&=&\frac{4f_T^2\dot{f_T}}{a\dot{\phi}^2(\dot{f_T}-3Hf_T)}~.
\eea
where we've made use of the relation: $\dot {f_T}=f_{TT}\dot T=-12f_{TT}H\dot{H}$. Note that different from $f(R)$ theories, the perturbation action contains also a higher-spatial-derivative term $\alpha_3(\partial^2\zeta)^2$. This is due to the fact that the constraint variable $\alpha$ now contains not only terms $\sim \dot\zeta$, but also terms $\sim \partial^2\zeta$, which is similar to the case of non-trivial kinetic coupling gravity theory explored in \cite{Qiu:2015aha}. Moreover, from action (\ref{pertaction}) one has the equation of motion of $\zeta$, or, the redefined variable $u\equiv\sqrt{\alpha_1}\zeta$, as
\be
\label{perteom}
u^{\prime\prime}+\left(\frac{\alpha_2}{\alpha_1}k^2-\frac{\alpha_3}{\alpha_1}k^4\right)u-\frac{z^{\prime\prime}}{z}u=0~.
\ee
In order to have the theory be free of ghost, one requires that $\alpha_1>0$ for all the time. Furthermore, from Eq. (\ref{perteom}), to eliminate the gradient instability in all region of $k$, one needs $\alpha_2>0$, $\alpha_3>0$ as well. Considering Eqs. (\ref{alpha1}), (\ref{alpha2}) and (\ref{alpha3}), one has the following conditions for stability:\\
\textbf{2) from $\alpha_1>0$},
\be
\label{con2}
\dot {f_T}\lessgtr 3Hf_T~{\rm for}~H\gtrless 0~, 
\ee
\textbf{3) from $\alpha_2>0$},
\be
\label{con3}
af_T+\frac{d}{dt}\left(\frac{3af_T^2}{\dot{f_T}-3Hf_T}\right)<0~,
\ee
\textbf{4) from $\alpha_3>0$},
\be
\label{con4}
\dot {f_T}\lessgtr 0~{\rm for}~H\gtrless 0~,
\ee
Note that actually Condition \textbf{2)} is contained in Condition \textbf{4)}. 

In the following, we analyze that under the conditions \textbf{1)} to \textbf{4)} obtained from the above subsections, what kind of $f(T)$ theory can survive. Interestingly, we find that, in order to obey all the conditions, one may have very harsh constraints on $f(T)$ theory. To be precise, we compile our results as a ``theorem'' which we think is useful for construction of bounce (or bounce inflation) scenarios in framework of $f(T)$ theories. The theorem is presented in the following:\\
{\it 1) the gravity theory with pure $f(T)$ cannot give rise to a bounce universe;\\
2) bounce can be realized with the help of exotic matter, e.g., a canonical scalar field. However, if the field doesn't contribute the perturbations, then in order for the perturbations be stable within all scales, at least at bounce point $\dot f_T$ cannot be a continuous function with respect to $t$.}

The first item is already demonstrated previously. To prove the second item, let's first introduce a lemma:\\
{\it Lemma. For any function $V(t)$ which satisfies $V(t)>0$ for $t>0$, $V(t)<0$ for $t<0$, or vice versa, then at $t=0$ point, $V(t)$ can either be vanishing, or become discontinuous.}

This is easy to prove. if $V(t)$ is continuous acrossing $t=0$ point, and assume $V(t=0)=V_\ast\neq0$, then we always have a small number $\varepsilon$, such that $V(t=0+\varepsilon)\approx V_\ast+V^\prime_\ast\varepsilon$, and $V(t=0-\varepsilon)\approx V_\ast-V^\prime_\ast\varepsilon$, where $V^\prime_\ast\varepsilon$ is the time derivative of $V(t)$ at $t=0$. So $V(t=0+\varepsilon)\cdot V(t=0-\varepsilon)=(V_\ast+V^\prime_\ast\varepsilon)(V_\ast-V^\prime_\ast\varepsilon)\approx V_\ast^2-V^{\prime 2}_\ast\varepsilon^2\approx V_\ast^2>0$ for small enough $\varepsilon$ and regular $V^\prime_\ast\varepsilon$. This violates the condition that $V(t)$ changes its sign before and after $t=0$ point. Proof completed. 

Now we prove the second item of the theorem. Since $H$ must change its sign when crossing the bounce point, in order to have $\alpha_1>0$, i.e., to eliminate the ghost problem, $\dot f_T-3Hf_T$ must change its sign when crossing the bounce point, which we set to be $t=0$. According to the lemma, $\dot f_T-3Hf_T$ can either cross 0 or become discontinuous. If $\dot f_T-3Hf_T$ crosses 0 at $t=0$, it means that $3af_T^2/(\dot f_T-3Hf_T)$ gets divergent when $t$ approaches to zero, unless$f_T$ also goes to zero to compensate the divergence. In that case, as $3af_T^2/(\dot f_T-3Hf_T)$ blows up, its time derivative will be at least positive, and $f_T$ is also positive considering the stability of tensor perturbations, which gives rise to $\alpha_2<0$, leading to a gradient instability.

The only loophole is to have $f_T$ also goes to zero at $t=0$, as mentioned before. However, it is also impossible. Since $f_T$ is constrained to be positive either before or after the bounce, $f_T\rightarrow 0$ means that $f_T$ decreases ($\dot f_T<0$) before the bounce, while increases ($\dot f_T>0$) after the bounce, contradicting with the requirement of $\alpha_3>0$. Therefore, the only way to have all the three $\alpha$'s be positive all the time is to have $\dot f_T-3Hf_T$ discontinuous, at least at $t=0$.

The discontinuity of $\dot f_T-3Hf_T$ implies that either $\dot f_T$ or $f_T$ be discontinuous, or both. However, the case where $f_T$ is discontinuous while $\dot f_T$ is not cannot be true. The reason is that according to the lemma and requirement of positivity of $\alpha_3$, $\dot f_T$ can either be discontinuous, or cross zero. If $\dot f_T\rightarrow 0$ at $t=0$, since $\dot f_T-3Hf_T$ cannot cross 0, it means $f_T$ has to be divergent (to compensate the vanishing of $H$), and moreover, $\dot f_T-3Hf_T\simeq -3Hf_T$. In this case, one has 
\be
\alpha_2\simeq\frac{4a\dot f_T}{H}+\frac{2a}{3H^2}(\ddot f_T-3\dot H f_T)~,\\
\ee
where one can see that, every term has a negative value. The first term is because $\dot f_T$ and $H$ must have opposite signs, the second term is because $\ddot f_T$ must be negative while $\dot f_T$ goes continuously from positive value to negative value, and the third term is because both $\dot H$ and $f_T$ are positive during bounce region. Therefore $\dot f_T$ must be discontinuous at least crossing the bounce point $t=0$. Hitherto the full proof completed. 

Actually, the implication of discontinuous function in cosmology solutions is not rare at all in the literature. For example, in Ref. \cite{Starobinsky:1992ts} people explore interesting observational effects brought by step-like functions in inflation model buildings. In next section, we will give an example of a bounce inflation scenario, which is modeled by $f(T)$ theory with discontinuous $\dot f_T$ at the bounce point.

{\it An concrete example.}
\label{example}
According to the theorem above, in this section, we provide a parameterized model of $f(T)$ theory that could have bounce inflation solution and satisfy all the requirements \textbf{1)} to \textbf{4)} for stabilities of perturbations. First of all, we parameterize the scale factor to be:
\be
a(t)=a_i|t-t_i|^{p_i}~,~i=1,2~{\rm for}~t</>0~,
\ee
where we assume that the bounce happens at $t=0$ point. And according to this, the Hubble parameter can be written as
$H=p_i/(t-t_i)$, which further leads to $T=-6H^2=-6p_i^2/(t-t_i)^2$.

Moreover, this parameterization will give the equation of state (EoS) $w_1=2/(3p_1)-1$ for $t<0$ and $w_2=2/(3p_2)-1$ for $t>0$. Considering the requirement on background demonstrated previously, one then has $0<p_1\leq1/3$ and $p_2\rightarrow +\infty$. For specific choice, we set $p_1=1/3$ and $p_2=100$.

%The Torsion scalar $T$ will be
%\be
%\label{Tpar}
%T=-6H^2=-6\left(\frac{p_i}{t-t_i}\right)^2~,
%\ee 

In order to solve $f(t)$ from Eq. (\ref{ft}), we also need to parameterize the scaling of matter density $\rho_m(t)$. For simplicity, we temporarily let 
\be
\label{rhopar}
\rho_m(t)=\rho_{mi}|t-t_i|^{2r_i}~,~i=1,2~{\rm for}~t</>0~,
\ee
so that we can write $\rho_{mi}\sim a^{\frac{2r_i}{p_i}}$ with $i=1,2$ for $t</>0$, which can give us the relation between the EoS of matter and that of background given by $f(T)$:  $w_{mi}=-2r_i/p_i-1=-r(1+w_i)-1$. Since both before and after bounce we have $w_i>-1$, therefore, if we use a single canonical scalar field as we did to describe the matter, such a relation requires that $r_i<0$ for both before and after the bounce. A more stringent constraint on $r_i$ will be given by the conditions \textbf{1)} to \textbf{4)}. They gives i) $r_i(1+2r_i)>0$, iii) $r_i(2r_i+3)/[(r_i+1)(2r_i+1)]<0$, ii) $-2r_i(1+r_i)/(1+2r_i)>0$, respectively, which can be combined to have $-3/2<r_i<-1$.

Considering all the parameterizations above as well as Eq. (\ref{ft}), 
%one can get
%\be
%\label{ftpar}
%f(t)=\left\{ \begin{array}{l} 
%\frac{2\rho_{m1}}{M_p^2(1+2r_1)}(t_1-t)^{2r_1}+\frac{\lambda_1}{t_1-t}~{\rm for}~t<0~,\\\\ 
%\frac{2\rho_{m2}}{M_p^2(1+2r_2)}(t-t_2)^{2r_2}-\frac{\lambda_2}{t-t_2}~{\rm for}~t>0~,\\ 
%\end{array}\right.
%\ee
one has the form of $f(T)$:
\be
\label{fT}
f(T)=\frac{2\rho_{mi}}{M_p^2(1+2r_i)}\left(\frac{T}{-6p_i^2}\right)^{2r_i}\pm\lambda_i\sqrt{\frac{T}{-6p_i^2}}~,
\ee
and $i=1,2$, $\pm=+,-$ for $t</>0$, respectively.

\begin{figure}[htbp]
\includegraphics[scale=0.4]{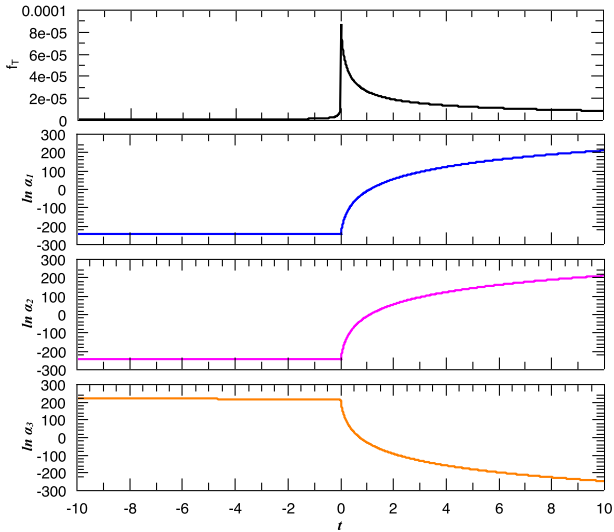}
\caption{From top to bottom are the plots of $f_T$, $\alpha_1$, $\alpha_2$ and $\alpha_3$ respectively, with respect to cosmic time $t$, while $t=0$ is the bounce point. Note that all these parameters are positive ($\alpha_i$'s are plotted in Logarithm scale.) while $f_T$ presents a sharp peak. }
\label{stability}
\end{figure}

Fig. \ref{stability} shows the time evolution of the parameters $\alpha_1$, $\alpha_2$, $\alpha_3$ as well as $f_T$ (derived from Eq. (\ref{fT})) which describes the stabilities of tensor and scalar perturbations. One can see that although $f_T$ appears continuous, it forms a sharp peak around the bounce point, demonstrating a discontinuity of its further derivative, $\dot f_T$ (In our numerical calculation, $\dot f_T(0_-)\approx 0.13M_p^3$ while $\dot f_T(0_+)\approx -4.39\times 10^{-4}M_p^3$). $\alpha_i$'s may also be discontinuous as they contain $\dot f_T$, nonetheless, all the parameters remain positive, leading to a totally stable bounce inflation solution. For parameter choices, we choose $a_1=10^{-100}$, $a_2=\sqrt[-1/3]{3}/10$, $t_1=10^{-3}/3M_p^{-1}$, $t_2=-0.1M_p^{-1}$, $p_1=1/3$, $p_2=100$, $\rho_{m1}=\sqrt[-5/2]{3}\times 10^{-5}M_p^4$, $\rho_{m2}=1M_p^4$, $\lambda_1=\lambda_2=0$, and $r_1=r_2=-5/4$, which means $w_{m1}=3/2$ for contracting phase with $w_1=1$, and $w_{m2}\approx -1$ for expanding phase with $w_2\approx 0$.These choices can ensure the continuity of $a(t)$, $T(t)$, $\rho_m(t)$ and $f(T)$, however, as all the degrees of freedom are thus used up, one has to abandon the continuities of further derivatives, namely $\dot f_T$. 

As a side remark, we note that scalar field with arbitrary EoS larger than $-1$ can be realized by potential parameterization, an example having been given in \cite{Liddle:1998xm}. In our case, the function form of the potential can be given by
$V(\phi)=V_{0i}[A_i(\phi-\phi_{0i})]^{2r_i/(r_i+1)}$, while $V_{0i}=\rho_{mi}[2+r(1+w_i)]/2$, $A_i=(1+r_i)/\sqrt{-r_i\rho_i(1+w_i)}$, and $\phi_0$ is a integral constant. 

{\it Conclusions.}
\label{conclusions}
In this letter, we investigated the properties of perturbations generated from $f(T)$ modified gravity theory applied to bounce inflation scenario. We calculated the perturbation action of $f(T)$ theory plus a scalar field, and found conditions for obtaining a stable bounce inflation solution. We found it is actually very difficult to satisfy all the conditions, and one way out is to give up the continuity of derivative function of $f(T)$, say, $\dot f_T$. An example of such a solution is also presented.

As a member of the big modified gravity family, $f(T)$ theory has many interesting features that are deserved further investigation. For example, as it breaks the local Lorentz symmetry \cite{Krssak:2015oua}, an interesting idea is to extend the current study to a more general torsion theory which restored the symmetry. One example of such a torsion theory, namely the Cartan theory, has been explored in Ref. \cite{Farnsworth:2017wzr}. Moreover, since we only consider perturbations from the gravity part. If the matter part also takes the role, the perturbation analysis will be more complicated since isotropic modes of perturbation also involved in. A higher order perturbations (Non-Gaussianities) of the system might also be interesting especially for future observational data. We leave all these topics for upcoming works.  

\begin{acknowledgments}
We thank Yi-Fu Cai, Jun Chen, Wenjie Hou, Keisuke Izumi, Ze Luan, Jiaming Shi, Taishi Katsuragawa, Emmanuel N. Saridakis, Yi-Peng Wu and Yun-Long Zheng for useful discussions. This work is supported by NSFC under Grant No: 11405069 and 11653002.
\end{acknowledgments}

\end{document}